\documentclass{article}

\usepackage{arxiv}

\usepackage[utf8]{inputenc} 
\usepackage[T1]{fontenc}    
\usepackage{hyperref}       
\usepackage{url}            
\usepackage{booktabs}       
\usepackage{amsfonts}       
\usepackage{nicefrac}       
\usepackage{microtype}      
\usepackage{lipsum}

\usepackage{amsmath,amsfonts,amssymb}
\usepackage{graphicx}
\usepackage{booktabs, siunitx, multirow, multicol}
\usepackage{wrapfig}
\usepackage{float}
\usepackage{xcolor}

\usepackage{newunicodechar}
\newunicodechar{ﬁ}{fi}
\newunicodechar{ﬀ}{ff}

\usepackage{mathtools}

\usepackage{pdfpages}

\title{Semi-supervised lung nodule retrieval}

\author{
  Mark Loyman \\
  Department of Biomedical-Engineering \\
  Tel-Aviv University\\
   \And
  Hayit Greenspan \\
  Department of Biomedical-Engineering \\
  Tel-Aviv University\\
}

\begin{document}
\maketitle

\begin{abstract}
Content based image retrieval (CBIR)  provides the clinician with visual information that can support, and hopefully improve, his or her decision making process.
Given an input query image, a CBIR system provides as its output a set of images, ranked by similarity to the query image. Retrieved images may come with relevant information, such as biopsy-based malignancy labeling, or categorization.
Ground truth on similarity between dataset elements (e.g. between nodules) is not readily available, thus greatly challenging  machine learning methods. 
Such annotations are particularly difficult to obtain, due to the subjective nature of the task, with high inter-observer variability requiring multiple expert annotators.
Consequently, past approaches have focused on manual feature extraction, while current approaches use auxiliary tasks, such as a binary classification task (e.g. malignancy), for which ground-true is more readily accessible. However, in a previous study, we have shown that binary auxiliary tasks are inferior to the usage of a rough similarity estimate that are derived from data annotations.  
The current study suggests a semi-supervised approach that involves two steps: 1) Automatic annotation of a given partially labeled dataset; 2) Learning a semantic similarity metric space based on the predicated annotations.
The proposed system is demonstrated in lung nodule retrieval using the LIDC dataset, and shows that it is feasible to learn embedding from predicted ratings. 
The semi-supervised approach has demonstrated a significantly higher discriminative ability than the fully-unsupervised reference.
\end{abstract}


\section{Introduction}

\subsection{Motivation}

In standard CADx, an automated system provides the clinician a diagnosis input, such as whether a given lesion is malignant or benign. The clinician then needs to integrate this input with his or her own diagnosis. One of the difficulties and limitations of this integration process is that there is very limited information given by the automated solution as to the reasoning behind the decision.
This explanatory gap remains a challenge when using  deep-learning methods.
A content based image retrieval (CBIR) system, often based on semantic image content, may assist the radiologist own decision making by providing the ability to review similar cases \cite{muramatsu2010,li2003investigation}. 

Li, Qiang, 2003 \cite{li2003investigation} used an observer study to demonstrate the benefit of CBIR in radiology. Five participating radiologists were given query patches of nodules, for which they were required to infer the likelihood of malignancy. This was performed twice: once with the query image alone and once with the aid of CBIR. In the second case they were shown 3 instances of the most similar malignant images and 3 instances of the most similar benign images. The average performance of the  five radiologists was shown to increase from 0.56 to 0.63 with the aid of similar nodules. The value of CBIR is however not limited to CADx, it is also an import tool in medical decision making for treatment and prognosis, by enabling access to the actual case data \cite{muller2017}.

\begin{figure}[ht]
\begin{minipage}[b]{.9\linewidth}
  \centering
  \centerline{\includegraphics[width=\linewidth]{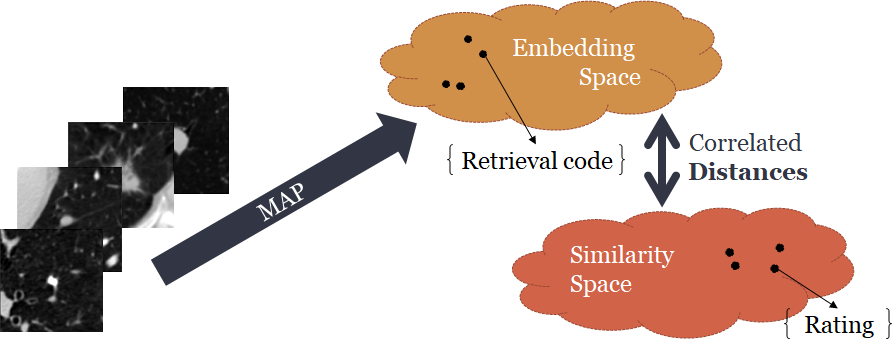}} 
\end{minipage}
\caption{
Semantic retrieval process: all database images are projected into the embedding space (each is essentially assigned a retrieval code). When a query image is presented, it is also projected into the embedding space, where nearest neighbours are retrieved. For the retrieval to be of semantic value, we require the distances in the embedding space to be correlated with the distances in the similarity space.
}
\label{fig:embedding}
\end{figure}

In the current work, we focus on providing a mapping from the image space to an embedding space: each image is mapped to a code that is used for its retrieval. We require  that the embedding spans a space, such that distances in the embedding space, preserve the semantic similarity of images and thus serve as  distances in the semantic space (see Fig. \ref{fig:embedding}).
Mapping into an embedding space, which is also referred to as a hash code, means that retrieval may be implemented using nearest neighbour searches. This enables the design of a scalable system that would implement retrieval over large scale distributed datasets. Liu et al 2007 \cite{liu2007clustering} present an approximate nearest neighbor search algorithm, which they apply to dataset of 1.5 billion images, spread over multiple machines. Their algorithm is designed to handle
generic embeddings, and is not restricted to any particular representation. 

\subsection{Dataset}
\label{sec:into-dataset}
\begin{figure}[ht]
\begin{minipage}[b]{.24\linewidth}
  \centering
  \centerline{\includegraphics[width=.9\linewidth]{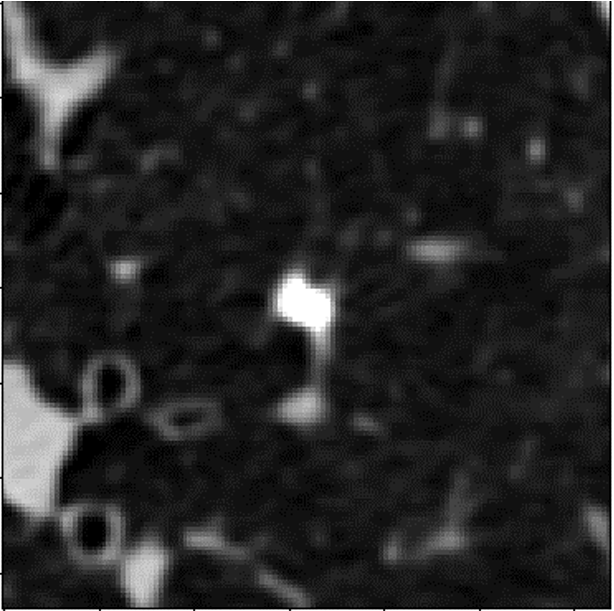}}
  \leftline{(a) Benign}
  \leftline{3, 1, 3, 3, 4, 3, 1, 5, 1}
  \medskip
\end{minipage}
\begin{minipage}[b]{.24\linewidth}
  \centering
  \centerline{\includegraphics[width=.9\linewidth]{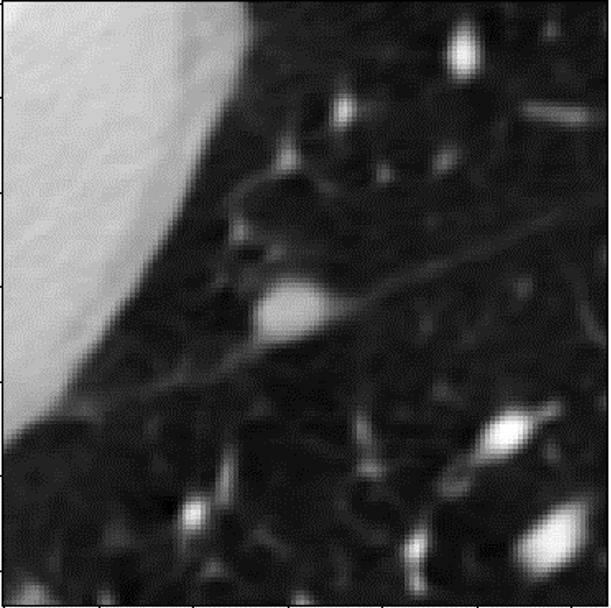}}
  \leftline{(b) Benign}
  \leftline{3, 1, 6, 5, 5, 1, 1, 5, 2}
  \medskip
\end{minipage}
\begin{minipage}[b]{.24\linewidth}
  \centering
  \centerline{\includegraphics[width=.9\linewidth]{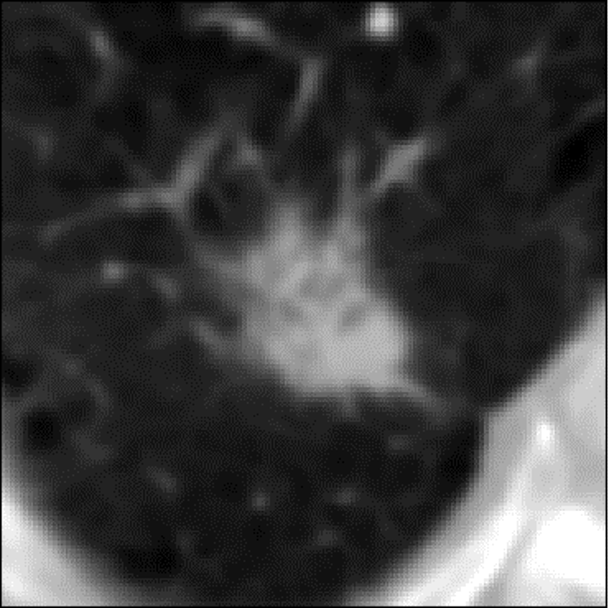}}
  \leftline{(c) Malignant}
  \leftline{5, 1, 6, 3, 3, 4, 1, 5, 5}
  \medskip
\end{minipage}
\hfill
\begin{minipage}[b]{.24\linewidth}
  \centering
  \centerline{\includegraphics[width=.9\linewidth]{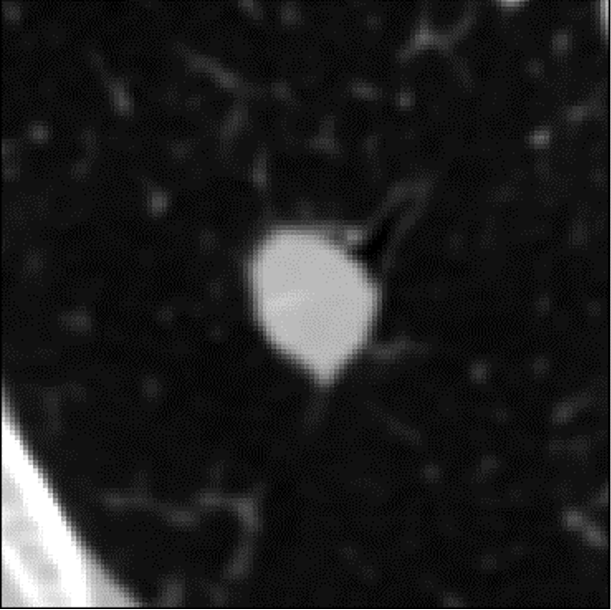}}
  \leftline{(d) Malignant}
  \leftline{4, 1, 6, 3, 4, 2, 2, 5, 4}
  \medskip
\end{minipage}
\caption{Examples of nodule patches. (a-b) benign (c-d) malignant. Ratings per nodule (mean of different raters)  specified below each nodule patch: Subtlety, Internal Structure, Calcification, Sphericity, Margin, Lobulation, Spiculation, Texture, Malignancy.}
\label{fig:nodules}
\end{figure}

The current study focuses on lung nodule retrieval from a database of CT scans, where patches taken from CT slices of pulmonary nodules are to be mapped into an embedding.
LIDC-IDRI \cite{lidc} is a dataset of thoracic CT scans of 1,010 patients. All the scans were annotated by up to four radiologists, where each one identified, segmented and evaluated separately the lung nodules of a diameter above $>3_{mm}$. Their evaluation also included ratings for a set of 9 characteristics: Subtlety, Internal structure, Calcification, Sphericity, Margin, Lobulation, Spiculation, Texture and Malignancy. The rating system was based on a discrete score of 1-5. 
Four examples of nodule patches are illustrated in Figure \ref{fig:nodules}: 2 benign (a, b) and 2 malignant (c, d). A rounded vector of the mean rating is bellow each nodule, with the characteristics ordered according to the listing above. The most prominent difference between a and b is the calcification: 3 (solid) and 6 (absent) accordingly. d compared to c has a more defined margin, is more lobulated, but less spiculated. 
The malignancy score is used to define malignancy classes: score of 1-2 is benign, score of 3 is unknown, and a score of 4-5 is malignant.

\subsection{Similarity}
\label{sec:intro-similarity}

Semantic image similarity lacks a clear and quantitative definition. This limitation requires manual acquisition of expert annotations to set the learning objective.
Li et al \cite{li2018muller} identifies two types of success criteria for CBIR: annotation-based and user-based. They advocate the user evaluation approach, putting forward two major drawback of the annotation approach: 1) the lack and difficulty in obtaining medical annotations, 2) annotations being too coarse for retrieval. 
However, the user-based approach remains highly dependant on manual expert feedback.
Unlike success criteria, in order to be used as a machine learning objective, generating pair-wise similarity ground-truth requires a quadratic amount of user evaluations.
Furthermore, observer studies have demonstrated high inter-observer variability, requiring the use of about 7-11 experts for similarity annotation of lung nodules\cite{li2003investigation}, mammography \cite{nishikawa2004,muramatsu2006}, and liver lesions\cite{faruque2015}. This implies an inherited limitation to similarity annotation and requiring that each nodule pair must be assessed by multiple experts.
This is a major bottleneck that prevents the assimilation of medical CBIR into an operational CADx work flow.

Li et al \cite{li2018muller}, in their review, only consider annotations that are based on a discrete set of classes. Indeed, recent studies adopted a degeneration of the similarity task into binary classification. Specifically, in the context of nodule retrieval, many recent CBIR studies focus on retrieval of nodules of the same binary malignancy class (malignant or benign) \cite{dhara2017,ferreira2017}. 
The success criteria that is used in these cases is based on a binary similarity measure: two nodules are similar if they belong to the same class. 

In our previous study \cite{loyman2019}, we demonstrated the effectiveness of using a similarity estimate that is based on an annotation vector of nodule characteristics, over a binary-classification approach. This supports the claim of Li et al \cite{li2018muller} that class-based annotations are not sufficient. At the same time we suggested an alternative approach that provides the necessary fine-grained similarity - to address their second limitation.
In the current study, we extend the above while  addressing the first limitation of the annotation-based approach: 
We present a semi-supervised methodology that would allow us to learn embeddings with limited similarity ground-truth, using automatically generated similarity labels.
Additionally, we introduce a new loss suitable for efficient metric-learning.

\subsection{Prior-Work}

The primary success metric we employ is the correlation of the embedding vector to the rating vector.
Kim 2010, \cite{kim2010} is the only study, we are aware of, that has used this metric. They used an artificial neural network (ANN) to learn a distance measure from a set of manual image features. Golden standard distance was based on the Ratings, with either Earth Mover's Distance  or Jeﬀrey's Divergence distances. Results are expressed as correlation of the learned distances to the golden standard rating distances. Their most successful attempt reached a correlation of only 0.129. 

Ferreira 2017 \cite{ferreira2017} used manual 3D features based on LIDC's segmentation data. Retrieval evaluation was conducted in a binary classification framework. Dhara 2017 \cite{dhara2017} performed automatic segmentation from a query point, as the basis for manual feature extraction. They used a combination of 3D and 2D features, taken from the largest slice. Both studies  used binary precision based on the malignancy class, to evaluate their work.

Radovanovic et al \cite{radovanovic2010} explore the behavior of points in high dimensional spaces. They introduce the phenomena of ``hubs": points that are close neighbors of many other points.
A strong bias exists to selection of such points during retrieval.   Similarly there are also orphan points that are not a neighbour of any query. This has been identified as a major obstacle for the retrieval with many studies suggesting a method to reduce the effect, most notably: Local Scaling\cite{zelnik2005self} and Mutual Proximity\cite{schnitzer2011mutualproximity} .
Furthermore, it has been shown that Hubness also occurs as the number of samples increases \cite{hara2015localized}. This makes it a measure of the potential scalability of a retrieval system, which is particularly critical when dealing with a medium sized dataset, such as the LIDC.
In a study by Li at el. \cite{li2003investigation} , involving retrieval of lung nodules, a case was presented where the same nodule appeared in the 6 nearest neighbors of 10 out of 20 queries, thus proving the practical significance of this measure. 
An extreme ``toy case" of a bad-hubness system is one that always returns one of two results: one malignant and one benign. Such a system may achieve perfect precision, but is clearly of no value in a retrieval task. 

In our previous study\cite{loyman2019}, we compared two approaches for metric learning: 1) implicit:  an auxiliary objective which is not similarity related and 2) explicit: similarity-based objective. 
The explicit approach was found to be inferior to the implicit approach. This was in contrast to the high potential that the explicit approach has due to it's direct optimization of the embedding space. This has led us to suspect that it's underperformance was due to our naive implementation. 
The explicit approach optimizes distances, therefore it is inherently bound to pair-wise evaluation of nodules. Consequently, due to the large quantity of possible pairs, the optimization process involves sampling of pairs each batch. If done randomly, this results in sub-optimal performance, as many pairs are trivial and can be considered as noise to the gradients.
Recent studies address this issue, and offer different adaptations for the loss and sampling process: lifted structured loss \cite{Song2016}, N-pair loss \cite{Sohn2016} and Cluster loss \cite{SongHO2017}.

However, all these losses are based on a definition of discrete set of classes, meaning, similarity is defined by class relation and is of binary nature: either similar: within-class, or different: inter-class. Therefore they are unsuitable for continuous distance metric learning: regression of distances. Consequently we present two novel batch-wise losses.

This paper is organized as follows: Section \ref{sec:Methods} describes the methods used to prepare the data, the architecture of the system that was implemented, as well as  the similarity measures that will be used. Section \ref{sec:Experiments} describes the experiments and Section \ref{sec:Results} presents their results. Section \ref{sec:Discussion} concludes with a discussion and summary of the results.
\section{Methods}
\label{sec:Methods}

\subsection{Dataset}
\label{sec:dataset}

Nodule extraction was done with the use of the pylidc tool, which was developed by Hancock et al.\cite{hancock2016}. Each annotation is given independently, so matching between the annotations is required to determine which annotations belong to the same physical nodule.
Nodule patches of size $64_{mm}\times64_{mm}$ were extracted from the 1,010 patients data and re-sampled to $0.5_{mm}$/px resolution using linear interpolation. Patches were scaled to HU window $[-300, 700]$ and then normalized.
Patches often contain multiple annotations, some of which belong to adjacent or overlapping nodules, and some to the same physical nodule but different rater.
In certain cases, a rater may have annotated a single non-uniform nodule, as two distinct nodules characterised with different parameters. 
Averaging the ratings over all annotations is trivial, yet loses much information, as well as simplifies the complexity we aim to capture in CBIR. A patch with several nodules needs to be similar to another patch with a similar collection of nodules, rather than a single nodule that is similar to the average nodule in the query patch. Therefore, each patch is assigned a set of ratings, and the similarity between patches is defined to be the distance between two sets of ratings. The distance will be described in Sec.\ref{sec:ratings}.

Patches are selected so that no nodule is sliced and to maximize the number of nodules they contain. For every nodule $n$, there are multiple annotations and 2D slices: slice $S_{n,\alpha,m}$ belongs to the $m$'th slice of annotation $\alpha$ in nodule $n$. For each annotation, slices are given weights according to their area relative to the maximal slice: $w_{n,\alpha,m} = \frac{area_{n,\alpha,m}}{max_{i \in 0..m}\{area_{n,\alpha,i}\}}$. Final weight per slice is given by summing over annotations: $w_{n,m} = \sum_{i \in 0..\alpha}\{w_{n,i,m}\}$. For each nodule, the slice with the maximal weight is selected as the patch for the dataset. Each nodule patch $S_n$ is assigned a set of $\alpha$ ratings and $\alpha$ weights.

Common practice \cite{ferreira2017,dhara2017,wei2018} is to discard patches with an ``unknown" malignancy rating. In contrast, we follow the conclusions of Loyman 2019 \cite{loyman2019} that in a retrieval context, all patches should be used. 

\subsection{LIDC ratings - semantic similarity}
\label{sec:ratings}

Retrieval performance evaluation requires a measure of similarity, therefore we follow Jabon et al \cite{jabon2009} and use LIDC’s ratings as a similarity code. Nonetheless, recent studies \cite{dhara2017,ferreira2017} often evaluate retrieval using a binary classification methodology. 
Several different metrics have been suggested for the evaluation of rating similarity in the LIDC dataset: Earth Movers Distance, Jeffery’s Divergence\cite{kim2010}, Jaccard and Cosine \cite{jabon2009} . 
Following the analysis in Loyman 2019 \cite{loyman2019} , we select the $L_2$ metric.

\begin{figure}[ht]
\begin{minipage}[b]{.9\linewidth}
  \centering
  \centerline{\includegraphics[width=.7\linewidth]{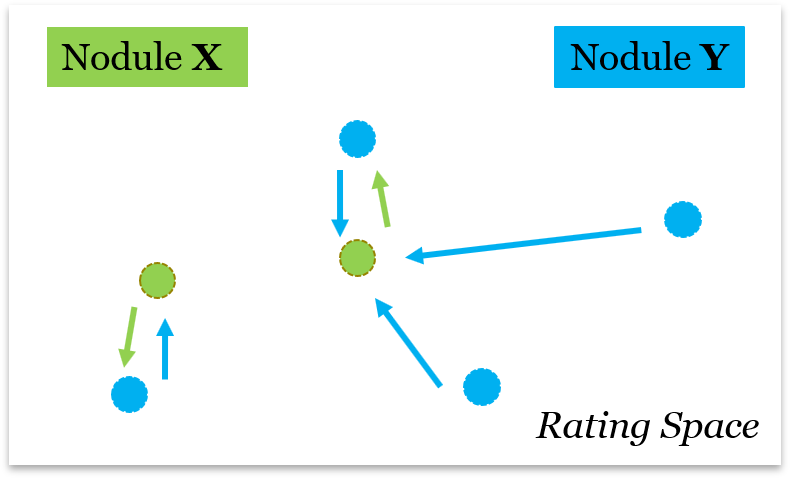}} 
\end{minipage}
\caption{
Each nodule is characterized by a set of rating vectors (points in rating-space). Similarity between two nodules is quantified by the distance between the two sets. To calculate the distance, we go over all ratings of nodule X and find their nearest rating in nodule Y. The distances of all pairs are averaged. This is then repeated for nodule Y.
}
\label{fig:rating-distance}
\end{figure}

As mentioned in Sec.\ref{sec:dataset}, our ground-truth semantic similarity between nodule patches is defined by the distance between two sets of ratings. 
Our distance is motivated by the hausdorff distance, which measures how far two sets in a metric space are from each other. This distance is calculated by finding pair-wise distances, for each point in one set to it's nearest neighbor in the other set. The final distance value is the maximum of all the above pair-wise distances. This has two related flaws: 1. it is very sensitive to outliers, 2. variation in any point in the set other than the extreme has no effect on the distance. We therefore modify the distance by taking the mean of the pair-wise distances. Our final distance can be expressed as (see Fig. \ref{fig:rating-distance}):

\begin{equation}
D(a,b) =  
    \frac{1}{2n} \sum_{i=1}^n min_{j \in [1, m]}  
        L_2(R_{a,i}, R_{b,j}) + 
    \frac{1}{2m}  \sum_{i=1}^m min_{j \in [1, n]} 
        L_2(R_{b,i}, R_{a,j})
\end{equation}

where $R_{x, y}$ is the rating vector of the $y$ elements in the set of nodule $x$, for patches a and b with n and m ratings in each set, and $L_2$ being the euclidean distance between two rating vectors. This can be intuitively interpreted as: for each rating in set a, take the mean of the nearest neighbors in set b.

\subsection{Architecture}
\label{sec:architecture}

\subsubsection{Network}

\begin{figure}[htp]
\begin{minipage}[b]{\linewidth}
(a) Rating regression network\newline
  \centerline{
  \includegraphics[width=.8\linewidth]{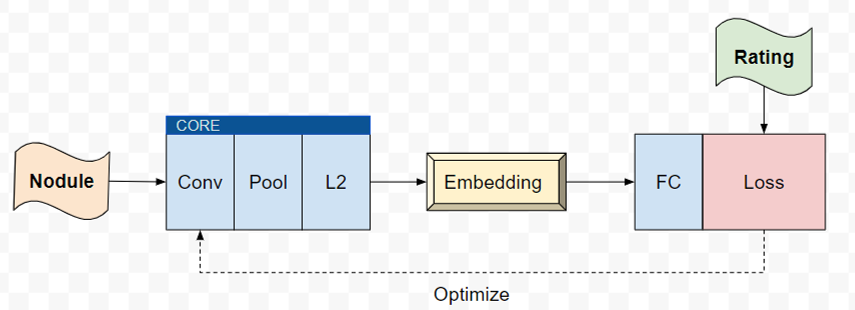}} 
\end{minipage}
\begin{minipage}[b]{\linewidth}
(b) Similarity network\newline
  \centerline{
  \includegraphics[width=.8\linewidth]{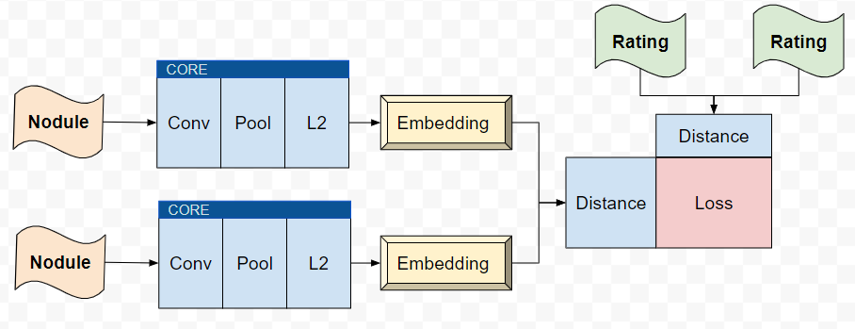}} 
\end{minipage}
\begin{minipage}[b]{\linewidth}
(c) Multi-task network\newline
  \centerline{
  \includegraphics[width=.8\linewidth]{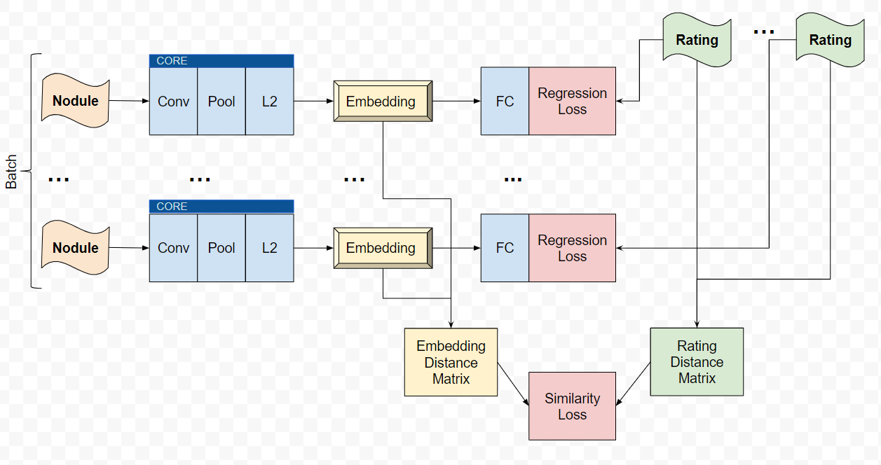}} 
\end{minipage}
%
\caption{Architecture block diagram. The core network includes a fully-convolutional sub-network, followed by a (max) pooling layer and L2 normalization - and is used for mapping the nodule patch to an embedding vector. Each embedding is fed into a FC (fully-connected) layer for rating regression using logcosh loss. All embedding within a batch are aggregated and fed to an L2-Distance-Matrix layer. The distance matrix is compared to the ratings-distance-matrix using one of the proposed similarity-losses.}
\label{fig:sys-arch}
\end{figure}

Our architectures use a core sub-network that maps nodule patches to an embedding vector of length 128\cite{loyman2019}. The core sub-network consists of a fully-convolutional part that is inspired by the Xception \cite{chollet2016} architecture, but adapted for smaller capacity to prevent overfiting. It is then sealed by Max-pooling and L2-normalization layers.
The ratings-regression network (Fig.\ref{fig:sys-arch}a) connects the embedding to a fully-connected layer of size 9 and with a linear activation. The output of the network is a vector of predicted ratings.
The similarity network (Fig.\ref{fig:sys-arch}b) uses the same Core, and applies an identical core to a pair of nodules, to create a pair of embeddings. An $L_2$ distance is then calculated, so the network outputs the embedding distance between the nodules. The objective for this network is the distance of the nodules in the ratings space, using the distance measure from section \ref{sec:ratings}. Such networks are typically based on the contrastive loss \cite{hadsell2006}.
The multi-task network (Fig.\ref{fig:sys-arch}c) combines regression loss: a local rating-regression loss that is applied per each individual nodule, and a similarity loss: a global loss that is calculated across the entire batch. An $L_2$-distance-matrix is calculated from the embedding, and from the ratings. Several similarity losses that use a distance-matrix as an objective are explored and introduced in the next sub-section.

\subsubsection{Similarity loss}

We define:  P is a $L_2$  distance matrix between predicted similarities; T an $L_2$ distance matrix of true similarities;  and B, the batch-size. We next introduce the following two losses:

1) Pearson correlation: 

\begin{equation}
    \sum_{b=1}^{B} \frac{B\sum_{i=1}^{B}T_{b,i}P_{b,i} - \sum_{i=1}^{B}T_{b,i}\sum_{i=1}^{B}P_{b,i}}{(B\sum_{i=1}^{B}T_{b,i}^2 - {\sum_{i=1}^{B}T_{b,i}}^2)(B\sum_{i=1}^{B}P_{b,i}^2 - {\sum_{i=1}^{B}P_{b,i}}^2)}
\end{equation}

The loss attempts to directly optimize the metric of interest (rating correlation). Additional variant is explored - Ranked correlation: as we don't care about the actual distances, but only in the relative distances, before applying pearson correlation, we normalize for each nodule, its distance to all other nodules, using a Softmax.

2) Kullback{\textendash}Leibler (KL) divergence:

\begin{equation}
\begin{multlined}
    \sum_{b=1}^{B}\sum_{i=1}^{B}\tilde{T}_{b,i}*Log(\frac{\tilde{T}_{b,i}}{\tilde{P}_{b,i}}) \\
    where \\
    \tilde{T}_{b,i}=
        \frac
            {{_{e}}^{T_{b,i}}}
            {\sum_{i=1}^{B}{_{e}}^{T_{b,i}}}, 
    \tilde{P}_{b,i}=
        \frac
            {{_{e}}^{P_{b,i}}}
            {\sum_{i=1}^{B}{_{e}}^{P_{b,i}}}
\end{multlined}    
\end{equation}

We consider each row of the distance-matrix as a discrete distribution of similarity to all nodules. We perform nodule-wise Softmax normalization (similar to the ranked-correlation) and use KL-divergence for the similarity quantification. KL-divergence measures the similarity (or dissimilarity) between two distributions, and can be seen as the amount of information lost, when using one distribution instead of the other.

\section{Experiments}
\label{sec:Experiments}

\subsection{Supervised-Metric-Learning}
\label{sec:exp-sup}

We evaluate two novel loss variants designed for a distance-matrix objective, and compare to a naive regression using logcosh-loss. Then we perform multi-task learning that employs both the implicit and explicit losses using two different schedule schemes: a) serial fine-tuning: first optimize the implicit loss and then the explicit loss. b) multi-task in 3 steps that are detailed in Table \ref{tab:schedule}.

\begin{table}[b]
\centering
\caption{Multi-task schedule scheme}
\begin{tabular}{@{}|c|cc|@{}}
\toprule
\rule[-1ex]{0pt}{3.5ex} Step & Regression loss & Similarity loss \\ \midrule
\rule[-1ex]{0pt}{3.5ex} 1 & 0.9 & 0.1 \\
\rule[-1ex]{0pt}{3.5ex} 2 & 0.5 & 0.5 \\
\rule[-1ex]{0pt}{3.5ex} 3 & 0.0 & 0.1 \\
\bottomrule
\end{tabular}
\label{tab:schedule}
\end{table}

The experiment is performed using 5-fold cross validation: 3/5 training, 1/5 validation and 1/5 test, with the subgroups described in Table \ref{tab:subsets}. The Validation set was used to select the length of training in epochs. Final results are evaluated over the Test set.

\begin{table}[b]
\centering
\caption{Data sub-sets}
\begin{tabular}{@{}|c|ccc|c|@{}}
\toprule
\rule[-1ex]{0pt}{3.5ex} Sub-set & Malignant & Benign & Unknown & Total \\ \midrule
\rule[-1ex]{0pt}{3.5ex} 0 & 114 & 237 & 211 & 562 \\
\rule[-1ex]{0pt}{3.5ex} 1 & 134 & 225 & 162 & 521 \\
\rule[-1ex]{0pt}{3.5ex} 2 & 105 & 232 & 202 & 539 \\
\rule[-1ex]{0pt}{3.5ex} 3 & 114 & 235 & 198 & 547 \\
\rule[-1ex]{0pt}{3.5ex} 4 & 118 & 229 & 196 & 543 \\
\bottomrule
\end{tabular}
\label{tab:subsets}
\end{table}

We evaluate the embedding using our golden-standard similarity measure: $L_2$ pearson correlation of the embedding-space to the rating-space, and the Hubness-Index. This is in contrast to the common practice of binary-class evaluation, which was shown to be non-discriminative of retrieval performance. This result is no surprise given that classification systems regard inner-class variability as noise, whereas retrieval systems must capture the rich variability.

Hubness \cite{radovanovic2010}: skewness of the distribution of k-occurrences $N_k$, which is defined as the number of times x occurs in the k nearest neighbors of other objects in the collection. For ease of evaluation, hubness-index was derived by normalizing the hubness to [0, 1] range, with 1 being no hubness, using: $\exp ^{-|hubness|}$. As Hubness is dependent on K, the results will be presented as mean of K=\{3, 5, 7, 11, 17\}.
To further understand the hubness phenomena we use reverse-nearest-neighbour search: we select the largest hub-nodule and display all queries that include that nodule in it's 2-nearest-neighbours results. This is presented in Fig. \ref{fig:hub-vis}.

\subsection{Unsupervised-Metric-Learning}
\label{sec:exp-unsup}

Data is split into 5 groups and we are evaluating 10 configurations, which are detailed in Table \ref{tab:crossvalidation}. The rating regression network is first trained to annotate the data. The predicted ratings annotations are then used to train a metric learning network. 
Step 1 - Rating predication: in each experiment (configuration), 2 groups are selected as the training for the rating-regression network, and 2 additional groups are used for validation. The validation sets are used to select the training duration (epoch). The 5$^{th}$ group was left out for the testing of retrieval. Predictions (ratings) are calculated for both the Validation and Test sets. 
Step 2 - A retrieval network is trained from the predicted ratings of the Validation set. Embeddings are evaluated for the Test set: the 5$^{th}$ group that was left out in previous step. For the retrieval objective, configurations {0, 1, 3, 4, 7} are used for validation and {2, 5, 6, 8, 9} are used for testing.
Step 3 - A retrieval network is trained from the true ratings of the validation set, in accordance with the 2$^{nd}$ step.

\begin{table}[t]
\centering
\caption{Cross-validation configurations: the usage of the 5 data sub-groups across the different steps. * Configurations \{0, 1, 3, 4, 7\} are used as Validation-set and \{2, 5, 6, 8, 9\} as Test-set}
\begin{center}
\begin{tabular}{@{}|c|ccc|cc|cc|@{}}
\toprule
\multirow{2}{*}{Configuration} & \multicolumn{3}{c}{Prediction} & \multicolumn{2}{c}{Supervised retrieval} & \multicolumn{2}{c}{Semi-supervised retrieval} \\
\rule[-1ex]{0pt}{3.5ex}  & Train. & Valid. & Test & Train. & Valid./Test* & Train. & Valid./Test* \\ 
\midrule
\rule[-1ex]{0pt}{3.5ex} 0 & 0, 1 & 2, 3 & 4 & 2, 3 & 4 & 2, 3 & 4\\
\rule[-1ex]{0pt}{3.5ex} 1 & 0, 2 & 1, 4 & 3 & 1, 4 & 3 & 1, 4 & 3\\
\rule[-1ex]{0pt}{3.5ex} 2* & 0, 3 & 1, 2 & 4 & 1, 2 & 4 & 1, 2 & 4\\
\rule[-1ex]{0pt}{3.5ex} 3 & 0, 4 & 1, 3 & 2 & 1, 3 & 2 & 1, 3 & 2\\
\rule[-1ex]{0pt}{3.5ex} 4 & 1, 2 & 3, 4 & 0 & 3, 4 & 0 & 3, 4 & 0\\
\rule[-1ex]{0pt}{3.5ex} 5* & 1, 3 & 2, 4 & 0 & 2, 4 & 0 & 2, 4 & 0\\
\rule[-1ex]{0pt}{3.5ex} 6* & 1, 4 & 0, 3 & 2 & 0, 3 & 2 & 0, 3 & 2\\
\rule[-1ex]{0pt}{3.5ex} 7 & 2, 3 & 0, 4 & 1 & 0, 4 & 1 & 0, 4 & 1\\
\rule[-1ex]{0pt}{3.5ex} 8* & 2, 4 & 0, 1 & 3 & 0, 1 & 3 & 0, 1 & 3\\
\rule[-1ex]{0pt}{3.5ex} 9* & 3, 4 & 0, 2 & 1 & 0, 2 & 1 & 0, 2 & 1\\
\bottomrule
\end{tabular}
\end{center}
\label{tab:crossvalidation}
\end{table}

The rating regression network is analyzed in detail and compared to the state of the art. 
Considering the inter-rater variability of the ratings, evaluation of the predicted ratings should be relative to the raters own uncertainty. 
We use the mahalanobis distance, which is a measure of how far a point is from a distribution, characterized by a covariance matrix. Specifically, we take the mahalanobis distance between the predicated rating vector and the set of true predication. Only nodule patches that have at least 4 ratings are considered.
Additionally, to get a better intuition for the quality of the results, we provide RMSE and correlation value for each rating parameter. Moreover, the following theoretic measures are calculated for : 1) An estimate for the regression potential is given by the inter-observer standard deviation, for each ROI that contains at least 4 ratings. This estimates the expert RMSE, as actual RMSE error for the human annotators could not be measured, due to the anonymity of the annotations. 2) An estimate for potential correlation is given by inter-observer correlation from study Ref.\cite{jabon2009}. 3) The entropy of the distribution of each of the parameters was calculated to quantify how informative each parameter is and identify degenerate cases. Current experiment differs from our previous work \cite{loyman2019} in the ratio  of the dataset that was available for training.

In the metric learning steps, we use the the retrieval network from Experiment \ref{sec:exp-sup}, and evaluate it using the rating correlation and hubness measures. Configurations {0,1,3,4,7} are used for Validation, to select the training schedule and training duration in epochs. Configurations {2,5,6,8,9} are used for Testing.

As a reference for the semi-supervised approach, we investigate a fully-unsupervised method. The last convolutional layer of the full xception network is passed through either max or average pooling, and $L_2$ normalized. This is used as a feature extractor to create embedding vectors, of length 2048, from the nodule patches.

\section{Results}
\label{sec:Results}

\subsection{Supervised metric learning}

Table \ref{tab:distance-matrix-loss} presents a comparison of the proposed distance-matrix loss functions for two variants of the core network: a) max-pooling, b) RMAC pooling. The naive Logcosh approach results in decent Hubness (0.51, 0.58), but very low rating-correlation (0.22, 0.26). Directly optimizing the pearson correlation gives improved rating-correlation (0.47, 0.41), but extremely low Hubness (0.1, 0.1). The ranked variant gives a more balanced result with correlation (0.45, 0.43) and Hubness (0.22, 0.25), however Hubness remains considerably low. The KL loss outperforms all in term of Hubness (0.7, 0.7), while maintaining a comparable correlation (0.42, 0.42).

\begin{figure}[ht]
\begin{minipage}[b]{.45\linewidth}
  \centering
  \centerline{\includegraphics[width=\linewidth]{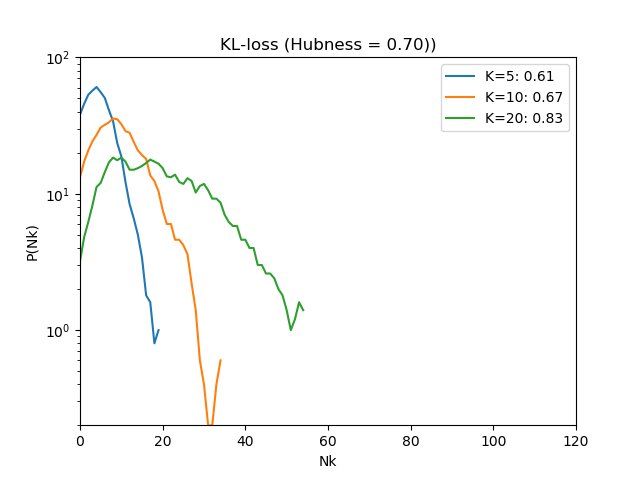}}
  \centerline{(a) k=7}
\end{minipage}
\begin{minipage}[b]{.45\linewidth}
  \centering
  \centerline{\includegraphics[width=\linewidth]{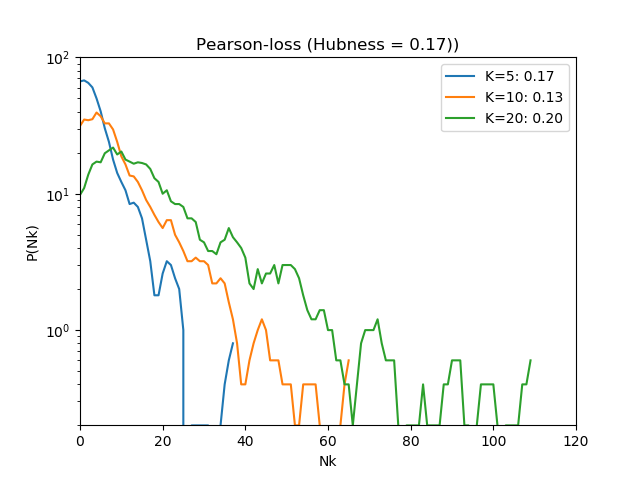}}
  \centerline{(a) k=11}
\end{minipage}
\caption{Rating space: histograms of k-occurrences (how many times a nodule appears as a nearest neighbor of other nodules). vertical axis is the (log) number of occurrences a of nodule within different k-neighborhoods. We can see that KL-loss is distributed around k, and a with a shorter tail.}
\label{fig:kocc}
\end{figure}

\begin{figure}[htp]
\begin{center}
\begin{minipage}[t]{.49\linewidth}
\centerline{(a) Pearson-loss Largest-Hub}
\leftline{
\fcolorbox{yellow}{yellow}{\includegraphics[width=64px]{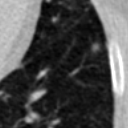}}
\includegraphics[width=64px]{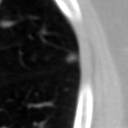}
\hspace{4pt}
\includegraphics[width=64px]{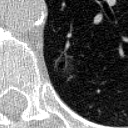}
}\leftline{
\includegraphics[width=64px]{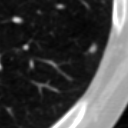} 
\hspace{4pt}
\includegraphics[width=64px]{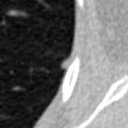}
\hspace{4pt}
\includegraphics[width=64px]{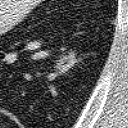}
}\leftline{
\includegraphics[width=64px]{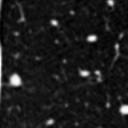}
\hspace{4pt}
\includegraphics[width=64px]{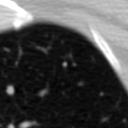}
\hspace{4pt}
\includegraphics[width=64px]{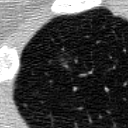}
}\leftline{
\includegraphics[width=64px]{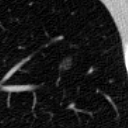}
\hspace{4pt}
\includegraphics[width=64px]{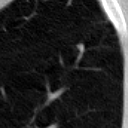}
\hspace{4pt}
\includegraphics[width=64px]{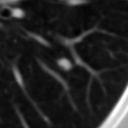}
}\leftline{
\includegraphics[width=64px]{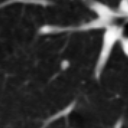}
\hspace{4pt}
\includegraphics[width=64px]{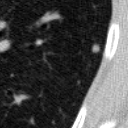}
\hspace{4pt}
\includegraphics[width=64px]{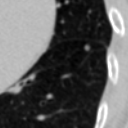}
}\leftline{
\includegraphics[width=64px]{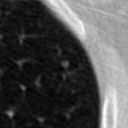}
\hspace{4pt}
\includegraphics[width=64px]{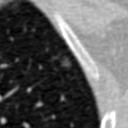}
\hspace{4pt}
\includegraphics[width=64px]{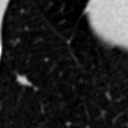}
}
\end{minipage}
\begin{minipage}[t]{.49\linewidth}
\centerline{(b) KL-loss Largest-Hub}
\rightline{
\fcolorbox{yellow}{yellow}{\includegraphics[width=64px]{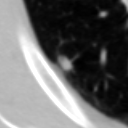}}
\includegraphics[width=64px]{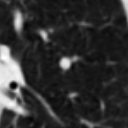}
\hspace{4pt}
\includegraphics[width=64px]{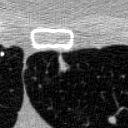}
}\rightline{
\includegraphics[width=64px]{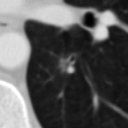}
\hspace{4pt}
\includegraphics[width=64px]{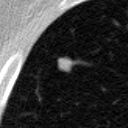}
\hspace{4pt}
\includegraphics[width=64px]{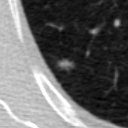}
}\rightline{
\includegraphics[width=64px]{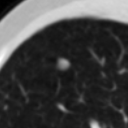}
\hspace{4pt}
\includegraphics[width=64px]{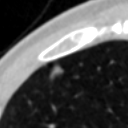}
\hspace{4pt}
\includegraphics[width=64px]{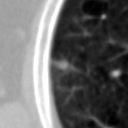}
}
\end{minipage}
\end{center}
\caption{Hubness visualization. The top-left image, also marked by yellow frame, is the largest hub in the systems embedding space. The rest of the images are the query-nodules that have the hub-nodule as their 2-nearest-neighbour. The large hubness (low hubness-index) that results from the pearson-loss is evident by the large hub (17 queries vs. 8 queries for the kl-loss) that it created in the embedding space. Full k-occurrences distribution is presented in Fig.\ref{fig:kocc}}
\label{fig:hub-vis}
\end{figure}

The k-occurrences distributions, for k=\{5, 10, 20\}, of the pearson and kl losses are given in Fig. \ref{fig:kocc}. Visualizations of the largest hubs of each loss, for k=\{2\}, are given in Fig. \ref{fig:hub-vis}. The larger hubness-index of the kl-loss (0.73 vs. 0.18) means less hubness - therefore we can see that it's largest hub is indeed smaller (8 vs. 17). Moreoever, better hubness index also means less orphans: nodules that can never be retrieved (94 vs. 145). 

KL-loss allows to better utilize the potential of the explicit approach, reaching top Hubness. However, correlation remains below the implicit rating-regression approach. We therefore combine both approaches in a multi-tasks training scheme. Results are in Table \ref{tab:multi-obj-training}. Multi-task training has allowed us to reach the correlation performance of the implicit rating-regression network (0.51), while improving the Hubness (from 0.59 to 0.79). A simpler two-step fine-tuning approach outperforms the single distance-matrix objective approach, but is overall below the multi-task option with 0.48 correlation and 0.75 Hubness.

\begin{table}[ht]
\centering
\caption{Comparison of Loss functions. Left (a): core network with max pooling; Right (b): core network with RMAC pooling. Both cases kl-loss reaches slightly less correlation that the correlation loss variants, however significantly improves the Hubness index.}
\begin{center}
\begin{tabular}{|l|cc|}
\toprule
(a) Loss & Correlation & Hubness \\
\midrule
distances-logcosh   & 0.22 & 0.51 \\
pearson-correlation & 0.47 & 0.10 \\
ranked-correlation  & 0.45 & 0.22 \\
kl                  & 0.42 & 0.70 \\
\bottomrule
\end{tabular}
\quad
\begin{tabular}{|l|cc|}
\toprule
(b) Loss & Correlation & Hubness \\
\midrule
distances-logcosh   & 0.26 & 0.58 \\
pearson-correlation & 0.41 & 0.10 \\
ranked-correlation  & 0.43 & 0.25 \\
kl                  & 0.42 & 0.70 \\
\bottomrule
\end{tabular}
\end{center}
\label{tab:distance-matrix-loss}
\end{table}

\begin{table}[ht]
\centering
\caption{Multi-task learning scheme combining rating-regression-logcosh-loss and distance-matrix-kl-loss.}
\begin{center}
\begin{tabular}{|ll|cc|}
\toprule
Objective & Loss & Correlation & Hubness \\
\midrule
Ratings             & Logcosh       & 0.51 & 0.59 \\
Siamese distance    & Contrastive   & 0.46 & 0.47 \\
Distance-matrix (DM)& kl            & 0.42 & 0.70 \\
Ratings \& DM       & Two-step finetune & 0.48 & 0.75 \\
Ratings \& DM       & Multi-task & 0.51 & 0.79 \\
\bottomrule
\end{tabular}
\end{center}
\label{tab:multi-obj-training}
\end{table}

\subsection{Unsupervised metric learning}

For the first rating-regression step, epoch 70 was selected based on the Validation set. 
The predicted ratings achieved mahalanobis distance of 1.4 from the raters distribution.
Full results on the Test set are in Table \ref{tab:rating-reg}. We can see that the rating-regression network, when trained on a smaller subset of 2/5 of the data, still gives a good RMSE and correlation. 
Compared to state-of-the-art, this network outperforms Li et al. \cite{li2017} in malignancy RMSE (0.72 vs. 0.89), but is below Tu et al. \cite{tu2017regression} (1.09 vs. 0.89) in texture RMSE. 
RMSE for Sphericity, Margin and Lobulation parameters was high relative to the inter-observer RMSE, but are in fact on par with the rest of the parameters.
This network was used to annotate the ratings for the unsupervised-training.

The supervised baseline was trained using Ground-Truth ratings with only 2/5 of the data, instead of 3/5 of the data, as was previously done. This resulted in a slight degradation of the results: rating-correlation was down to 0.46 (from 0.51) and hubness-index was down to 0.72 (from 0.79).
Semi-supervised training using predicated ratings, resulted in a correlation degradation of 8.7\%, whereas hubness has increased by 5.2\%. 
This is a drastic improvement over the the fully unsupervised alternative that resulted in a correlation degradation of 78\% and hubness degradation of 69\%. Final comparisons on the Test-data are summarized in Table \ref{tab:result-semisupervised}. The results for the unsupervised approach are presented in Table \ref{tab:unsupervised}.

\begin{table}[ht]
\centering
\caption{Rating-Regression evaluation}
\begin{center}
\begin{tabular}{@{}|l|cc|cc|c|@{}}
\toprule
\multirow{2}{*}{Parameter} & \multicolumn{2}{c}{RMSE} & \multicolumn{2}{c}{Correlation}  & \multirow{2}{*}{Entropy} \\
 & Our & Inter-Observer & Our & Inter-Observer \\ 
\midrule
\rule[-1ex]{0pt}{3.5ex} Subtlety            & 0.97 & 0.84 & 0.50 & 0.76 & 0.84 \\
\rule[-1ex]{0pt}{3.5ex} Internal Structure  & 1.10 & 0.80 & NA   & NA   & 0.03 \\
\rule[-1ex]{0pt}{3.5ex} Calcification       & 0.84 & 0.39 & NA   & NA   & 0.26 \\
\rule[-1ex]{0pt}{3.5ex} Sphericity          & 0.84 & 0.47 & 0.25 & 0.59 & 0.79 \\
\rule[-1ex]{0pt}{3.5ex} Margin              & 0.97 & 0.37 & 0.57 & 0.74 & 0.90 \\
\rule[-1ex]{0pt}{3.5ex} Lobulation          & 0.90 & 0.27 & 0.44 & 0.58 & 0.70 \\
\rule[-1ex]{0pt}{3.5ex} Spiculation         & 0.95 & 0.84 & 0.47 & 0.62 & 0.65 \\
\rule[-1ex]{0pt}{3.5ex} Texture             & 1.09 & 0.84 & 0.64 & 0.74 & 0.61 \\
\rule[-1ex]{0pt}{3.5ex} Malignancy          & 0.72 & 0.84 & 0.69 & 0.65 & 0.94 \\
\bottomrule
\end{tabular}
\end{center}
\label{tab:rating-reg}
\end{table}

\begin{table}[b]
\centering
\caption{Unsupervised retrieval using the xception network}
\begin{tabular}{@{}|cc|cc|@{}}
\toprule
\rule[-1ex]{0pt}{3.5ex} Weight initialization & pooling & Correlation & Hubness \\ \midrule
\rule[-1ex]{0pt}{3.5ex} random init & avg & 0.02 & 0.74 \\
\rule[-1ex]{0pt}{3.5ex} imagenet    & avg & 0.1 & 0.24 \\
\rule[-1ex]{0pt}{3.5ex} imagenet    & max & 0.1 & 0.21 \\
\bottomrule
\end{tabular}
\label{tab:unsupervised}
\end{table} .

\begin{table}[ht]
\centering
\caption{Supervised vs. Semi-supervised. All variants use multi-task learning. (*) training using 3/5 of the data}
\begin{center}
\begin{tabular}{|l|cc|}
\toprule
Method & Correlation & Hubness \\
\midrule
Supervised*     & 0.51 & 0.79 \\
Supervised      & 0.46 & 0.77 \\
Semi-supervised & 0.42 & 0.81 \\
Unsupervised    & 0.10 & 0.24 \\
\midrule
Semi-supervised cost    & -8.7\%    & +5.2\%    \\
Unsupervised cost       & -78\%     & -69\%      \\
\bottomrule
\end{tabular}
\end{center}
\label{tab:result-semisupervised}
\end{table}

\begin{figure}[htp]
\begin{center}
\centerline{
\begin{minipage}[]{.21\linewidth}
    \begin{center}
    Query
    \end{center}
\end{minipage}
\begin{minipage}[]{.35\linewidth}
    \begin{center}
    Retrieval by semi-supervised CBIR
    \end{center}
\end{minipage}
\begin{minipage}[]{.35\linewidth}
    \begin{center}
    Retrieval by Ratings
    \end{center}
\end{minipage}
}
\centerline{
%
\begin{minipage}[]{.21\linewidth}
    (a)
    \fcolorbox{blue}{blue!60}{
    \includegraphics[width=64px]{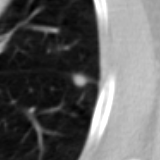}}
\end{minipage}
\begin{minipage}[]{.35\linewidth}
    \centerline{
    \fcolorbox{blue}{blue!60}{
    \includegraphics[width=64px]{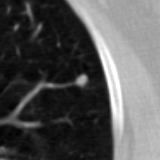}}
    \fcolorbox{blue}{blue!60}{
    \includegraphics[width=64px]{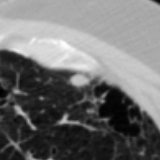}}
    } \centerline{
    \fcolorbox{blue}{blue!60}{
    \includegraphics[width=64px]{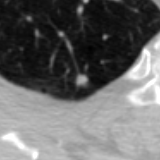}}
    \fcolorbox{gray}{gray!60}{
    \includegraphics[width=64px]{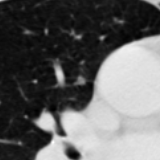}}
    }
\end{minipage}
\begin{minipage}[]{.35\linewidth}
    \centerline{
    \fcolorbox{blue}{blue!60}{
    \includegraphics[width=64px]{images/patches/T4-016.png}}
    \fcolorbox{blue}{blue!60}{
    \includegraphics[width=64px]{images/patches/T4-088.png}}
    } \centerline{
    \fcolorbox{gray}{gray!60}{
    \includegraphics[width=64px]{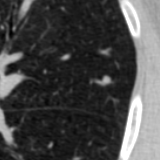}}
    \fcolorbox{blue}{blue!60}{
    \includegraphics[width=64px]{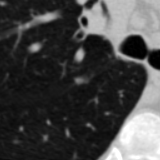}}
    }
\end{minipage}
}
\vspace{6pt}
%
%
\centerline{
\begin{minipage}[]{.21\linewidth}
    (b)
    \fcolorbox{red}{red!60}{
    \includegraphics[width=64px]{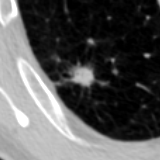}}
\end{minipage}
\begin{minipage}[]{.35\linewidth}
    \centerline{
    \fcolorbox{red}{red!60}{
    \includegraphics[width=64px]{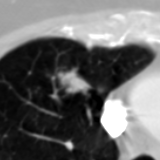}}
    \fcolorbox{red}{red!60}{
    \includegraphics[width=64px]{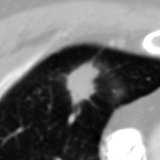}}
    } \centerline{
    \fcolorbox{red}{red!60}{
    \includegraphics[width=64px]{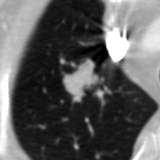}}
    \fcolorbox{red}{red!60}{
    \includegraphics[width=64px]{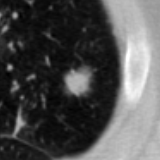}}
    }
\end{minipage}
\begin{minipage}[]{.35\linewidth}
    \centerline{
    \fcolorbox{red}{red!60}{
    \includegraphics[width=64px]{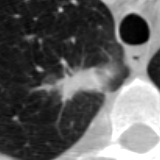}}
    \fcolorbox{red}{red!60}{
    \includegraphics[width=64px]{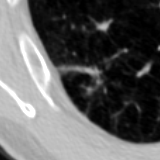}}
    } \centerline{
    \fcolorbox{red}{red!60}{
    \includegraphics[width=64px]{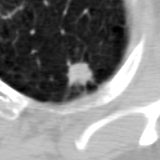}}
    \fcolorbox{red}{red!60}{
    \includegraphics[width=64px]{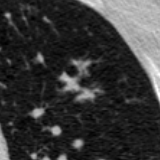}}
    }
\end{minipage}
}
\vspace{6pt}
\centerline{
%
\begin{minipage}[]{.21\linewidth}
    (c)
    \fcolorbox{gray}{gray!60}{
    \includegraphics[width=64px]{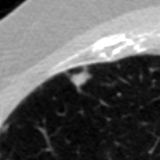}}
\end{minipage}
\begin{minipage}[]{.35\linewidth}
    \centerline{
    \fcolorbox{blue}{blue!60}{
    \includegraphics[width=64px]{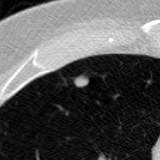}}
    \fcolorbox{gray}{gray!60}{
    \includegraphics[width=64px]{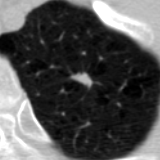}}
    } \centerline{
    \fcolorbox{gray}{gray!60}{
    \includegraphics[width=64px]{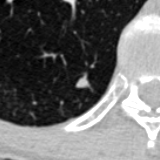}}
    \fcolorbox{red}{red!60}{
    \includegraphics[width=64px]{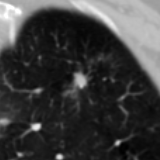}}
    }
\end{minipage}
\begin{minipage}[]{.35\linewidth}
    \centerline{
    \fcolorbox{gray}{gray!60}{
    \includegraphics[width=64px]{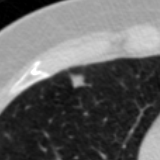}}
    \fcolorbox{gray}{gray!60}{
    \includegraphics[width=64px]{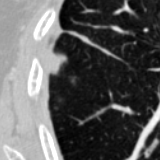}}
    } \centerline{
    \fcolorbox{gray}{gray!60}{
    \includegraphics[width=64px]{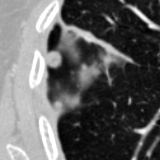}}
    \fcolorbox{gray}{gray!60}{
    \includegraphics[width=64px]{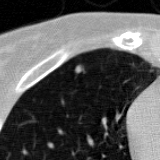}}
    }
\end{minipage}
}
\end{center}
\caption{
Semi-supervised CBIR vs Rating based: Two retrieval examples for (a) benign query, (b) malignant query and (c) unknown query. 
Left: query nodule patch, Center: top 4 retrieval results for the semi-supervised CBIR system, Right: top 4 retrieval by ground-truth ratings. Benign patches indicated by blue frame; Malignant patches indicated by red frame; Unknown patches indicated by gray frame. 
(a) The CBIR results include two patches that are also in the top retrieval by ratings. Moreover, these are the two which are most similar to the query. 
(b) Although none of the top-rating results are retrieved by our system, it is evident that the CBIR results bear more visual resemblance to the query. 
(c) For an unknown query, we received both a benign result and a malignant results, both similar to the query. Such results can assist the clinician with a differential diagnosis.}
\label{fig:ret-cbir-rbir}
\end{figure}

Examples of nodule retrieval are presented in Fig. \ref{fig:ret-cbir-rbir}.

\section{Discussion}
\label{sec:Discussion}

Rating regression: Seeing as the rating regression network was chosen for the retrieval task, it might be tempting to use the ratings themselves as the basis for retrieval. Such approach would result in a strong coupling of the retrieval process on a particular representation of similarity. Moreover, using the predicted ratings can not capture some finer visual content similarity concepts, as is captured by the embedding. This is demonstrated in Fig. \ref{fig:ret-cbir-rbir}.

Supervised metric learning: The distance-matrix loss, in particular the pearson loss, gives sub-optimal correlation results. This may be surprising, as we are directly optimization our evaluation metric. This can be explained by the batch size dependency. Normally, batch size is a hyper-parameter that is tuned to improve the optimisation process via mini-batch methodology \cite{masters2018revisiting}. In the case of pearson-loss, the batch size also effects the accuracy of the pearson approximation. In terms of pearson approximation, we want the batch size to be as large as possible, however we are not ``free" to increase it as we wish, due to its strong effect on optimisation.

Semi-supervised metric learning: We have demonstrated the feasibility of using automatic-annotations as an objective for metric learning, thereby opening the door for semi-supervised CBIR in the medical domain. While we see an expected decline in rating-correlation. The hubness-index is actually improved. We suspect that this is due to the fact that the rating labels in the semi-supervised task are much cleaner and have a descriptive capacity that is appropriate for the used network. 
The utilization of partial similarity labels has improved both correlation and hubness, relative to the unsupervised alternative.

The methodology presented in this study is not limited to the LIDC and its rating system. Any dataset can be used, as long as the following conditions hold: 1) the dataset either is manually annotated or it is possible to generate automatic annotations for it, 2) the annotations needs to be such that would allow to define a continuous similarity distance measure. RadLex is a popular ontology in the medical domain, and the automatic generation of RadLex terms is well studied \cite{annotation2019, hassanpour2016information, shah2009ontology}. Moreover, recent works have suggested to use it as a basis for CBIR, and have consequently defined a similarity measure between two sets of RadLex terms \cite{spanier2017radlex}. This means that any radiology dataset can be annotated using RadLex and used to build a CBIR system.

In future work we plan to explore the following: 
Integration of hubness-reduction techniques, such as Local Scaling\cite{zelnik2005self} and Mutual Proximity\cite{schnitzer2011mutualproximity} , into the similarity loss. 
Moreover, we plan to further validate our methods on additional datasets. The semi-supervised methodology should be applied to an additional lung-nodule dataset, using a rating-regression network that is learnt from LIDC. 
The proposed similarity loss, should be tested on a dataset from a different domain.
To conclude, in the current work we have shown that it is possible to learn an embedding that is correlated to a desired semantic label, when only a subset of the data is labeled. This can be generalized to any domain, as the methodology and presented loss function, do not make any assumptions regarding the labels. Any labeling system that has a semantic meaning and allows to define a similarity score between two items are valid. In the medical domain, RadLex annotations are a natural candidate. Once proven on additional cases, this can advance the design of real world large scale retrieval systems.

\section{Disclosures}
The authors declare no conflict of interest.
\section{Acknowledgement}

The authors acknowledge the National Cancer Institute and the Foundation for the National Institutes of Health, and their critical role in the creation of the free publicly available LIDC/IDRI Database used in this study.

\bibliographystyle{unsrt}  

\bibliography{references}  

\begin{thebibliography}{10}

\bibitem{muramatsu2010}
Chisako Muramatsu, Robert~A Schmidt, Junji Shiraishi, Qiang Li, and Kunio Doi.
\newblock Presentation of similar images as a reference for distinction between
  benign and malignant masses on mammograms: analysis of initial observer
  study.
\newblock {\em Journal of digital imaging}, 23(5):592--602, 2010.

\bibitem{li2003investigation}
Qiang Li, Feng Li, Junji Shiraishi, Shigehiko Katsuragawa, Shusuke Sone, et~al.
\newblock Investigation of new psychophysical measures for evaluation of
  similar images on thoracic computed tomography for distinction between benign
  and malignant nodules.
\newblock {\em Medical Physics}, 30(10):2584--2593, 2003.

\bibitem{muller2017}
Henning M{\"u}ller and Devrim Unay.
\newblock Retrieval from and understanding of large-scale multi-modal medical
  datasets: A review.
\newblock {\em IEEE Transactions on Multimedia}, 19(9):2093--2104, 2017.

\bibitem{liu2007clustering}
Ting Liu, Charles Rosenberg, and Henry~A Rowley.
\newblock Clustering billions of images with large scale nearest neighbor
  search.
\newblock In {\em 2007 IEEE Workshop on Applications of Computer Vision
  (WACV'07)}, pages 28--28. IEEE, 2007.

\bibitem{lidc}
Samuel~G Armato~III, Geoffrey McLennan, Luc Bidaut, Michael~F. McNitt-Gray, and
  et~al.
\newblock Data from lidc-idri. the cancer imaging archive., 2015.

\bibitem{li2018muller}
Zhongyu Li, Xiaofan Zhang, Henning M{\"u}ller, and Shaoting Zhang.
\newblock Large-scale retrieval for medical image analytics: A comprehensive
  review.
\newblock {\em Medical image analysis}, 43:66--84, 2018.

\bibitem{nishikawa2004}
Robert~M Nishikawa, Yongyi Yang, Dezheng Huo, Miles Wernick, Charlene~A
  Sennett, John Papaioannou, and Liyang Wei.
\newblock Observers' ability to judge the similarity of clustered
  calcifications on mammograms.
\newblock In {\em Medical Imaging 2004: Image Perception, Observer Performance,
  and Technology Assessment}, volume 5372, pages 192--199. International
  Society for Optics and Photonics, 2004.

\bibitem{muramatsu2006}
Chisako Muramatsu, Qiang Li, Robert Schmidt, Kenji Suzuki, Junji Shiraishi,
  Gillian Newstead, et~al.
\newblock Experimental determination of subjective similarity for pairs of
  clustered microcalcifications on mammograms: observer study results.
\newblock {\em Medical Physics}, 33(9):3460--3468, 2006.

\bibitem{faruque2015}
Jessica Faruque, Christopher~F Beaulieu, Jarrett Rosenberg, Daniel Rubin,
  Dorcas Yao, and Sandy Napel.
\newblock Content-based image retrieval in radiology: analysis of variability
  in human perception of similarity.
\newblock {\em Journal of Medical Imaging}, 2(2):025501, 2015.

\bibitem{dhara2017}
Ashis~Kumar Dhara, Sudipta Mukhopadhyay, Anirvan Dutta, Mandeep Garg, and
  Niranjan Khandelwal.
\newblock Content-based image retrieval system for pulmonary nodules: Assisting
  radiologists in self-learning and diagnosis of lung cancer.
\newblock {\em Journal of digital imaging}, 30(1):63--77, 2017.

\bibitem{ferreira2017}
Jos{\'e}~Raniery Ferreira, Paulo~Mazzoncini de~Azevedo-Marques, and
  Marcelo~Costa Oliveira.
\newblock Selecting relevant 3d image features of margin sharpness and texture
  for lung nodule retrieval.
\newblock {\em International journal of computer assisted radiology and
  surgery}, 12(3):509--517, 2017.

\bibitem{loyman2019}
Mark Loyman and Hayit Greenspan.
\newblock Lung nodule retrieval using semantic similarity estimates.
\newblock In {\em Medical Imaging 2019: Computer-Aided Diagnosis}, volume
  10950. International Society for Optics and Photonics, 2019.

\bibitem{kim2010}
Robert Kim, Grace Dasovich, Runa Bhaumik, Richard Brock, Jacob~D Furst, and
  Daniela~S Raicu.
\newblock An investigation into the relationship between semantic and content
  based similarity using lidc.
\newblock In {\em Proceedings of the international conference on Multimedia
  information retrieval}, pages 185--192. ACM, 2010.

\bibitem{radovanovic2010}
Milo{\v{s}} Radovanovi{\'c}, Alexandros Nanopoulos, and Mirjana Ivanovi{\'c}.
\newblock Hubs in space: Popular nearest neighbors in high-dimensional data.
\newblock {\em Journal of Machine Learning Research}, 11(Sep):2487--2531, 2010.

\bibitem{zelnik2005self}
Lihi Zelnik-Manor and Pietro Perona.
\newblock Self-tuning spectral clustering.
\newblock In {\em Advances in neural information processing systems}, pages
  1601--1608, 2005.

\bibitem{schnitzer2011mutualproximity}
Dominik Schnitzer, Arthur Flexer, Markus Schedl, and Gerhard Widmer.
\newblock Using mutual proximity to improve content-based audio similarity.
\newblock In {\em ISMIR}, volume~11, pages 79--84, 2011.

\bibitem{hara2015localized}
Kazuo Hara, Ikumi Suzuki, Masashi Shimbo, Kei Kobayashi, Kenji Fukumizu, and
  Milo{\v{s}} Radovanovi{\'c}.
\newblock Localized centering: Reducing hubness in large-sample data.
\newblock In {\em Twenty-Ninth AAAI Conference on Artificial Intelligence},
  2015.

\bibitem{Song2016}
Hyun~Oh Song, Yu~Xiang, Stefanie Jegelka, and Silvio Savarese.
\newblock Deep metric learning via lifted structured feature embedding.
\newblock {\em 2016 IEEE Conference on Computer Vision and Pattern Recognition
  (CVPR)}, pages 4004--4012, 2016.

\bibitem{Sohn2016}
Kihyuk Sohn.
\newblock Improved deep metric learning with multi-class n-pair loss objective.
\newblock In {\em NIPS}, 2016.

\bibitem{SongHO2017}
H.~O. {Song}, S.~{Jegelka}, V.~{Rathod}, and K.~{Murphy}.
\newblock Deep metric learning via facility location.
\newblock In {\em 2017 IEEE Conference on Computer Vision and Pattern
  Recognition (CVPR)}, pages 2206--2214, July 2017.

\bibitem{hancock2016}
Matthew~C Hancock and Jerry~F Magnan.
\newblock Lung nodule malignancy classification using only
  radiologist-quantified image features as inputs to statistical learning
  algorithms: probing the lung image database consortium dataset with two
  statistical learning methods.
\newblock {\em Journal of Medical Imaging}, 3(4):044504, 2016.

\bibitem{wei2018}
Guohui Wei, Hui Cao, He~Ma, Shouliang Qi, Wei Qian, and Zhiqing Ma.
\newblock Content-based image retrieval for lung nodule classification using
  texture features and learned distance metric.
\newblock {\em Journal of medical systems}, 42(1):13, 2018.

\bibitem{jabon2009}
Sarah~A Jabon, Daniela~S Raicu, and Jacob~D Furst.
\newblock Content-based versus semantic-based retrieval: an lidc case study.
\newblock In {\em Medical Imaging 2009: Image Perception, Observer Performance,
  and Technology Assessment}, volume 7263, page 72631L. International Society
  for Optics and Photonics, 2009.

\bibitem{chollet2016}
Fran{\c{c}}ois Chollet.
\newblock Xception: Deep learning with depthwise separable convolutions.
\newblock {\em arXiv preprint}, 2016.

\bibitem{hadsell2006}
Raia Hadsell, Sumit Chopra, and Yann LeCun.
\newblock Dimensionality reduction by learning an invariant mapping.
\newblock In {\em Computer vision and pattern recognition, 2006 IEEE computer
  society conference on}, volume~2, pages 1735--1742. IEEE, 2006.

\bibitem{li2017}
Xiuli Li, Yueying Kao, Wei Shen, Xiang Li, and Guotong Xie.
\newblock Lung nodule malignancy prediction using multi-task convolutional
  neural network.
\newblock In {\em Medical Imaging 2017: Computer-Aided Diagnosis}, volume
  10134, page 1013424. International Society for Optics and Photonics, 2017.

\bibitem{tu2017regression}
Xiaoguang Tu, Mei Xie, Jingjing Gao, Zheng Ma, Daiqiang Chen, Qingfeng Wang,
  Samuel~G Finlayson, Yangming Ou, and Jie-Zhi Cheng.
\newblock Automatic categorization and scoring of solid, part-solid and
  non-solid pulmonary nodules in ct images with convolutional neural network.
\newblock {\em Scientific reports}, 7(1):8533, 2017.

\bibitem{masters2018revisiting}
Dominic Masters and Carlo Luschi.
\newblock Revisiting small batch training for deep neural networks.
\newblock {\em arXiv preprint arXiv:1804.07612}, 2018.

\bibitem{annotation2019}
K.~{Yan}, Y.~{Peng}, Z.~{Lu}, and R.~M. {Summers}.
\newblock Fine-grained lesion annotation in ct images with knowledge mined from
  radiology reports.
\newblock In {\em 2019 IEEE 16th International Symposium on Biomedical Imaging
  (ISBI 2019)}, pages 285--288, April 2019.

\bibitem{hassanpour2016information}
Saeed Hassanpour and Curtis~P Langlotz.
\newblock Information extraction from multi-institutional radiology reports.
\newblock {\em Artificial intelligence in medicine}, 66:29--39, 2016.

\bibitem{shah2009ontology}
Nigam~H Shah, Clement Jonquet, Annie~P Chiang, Atul~J Butte, Rong Chen, and
  Mark~A Musen.
\newblock Ontology-driven indexing of public datasets for translational
  bioinformatics.
\newblock In {\em BMC bioinformatics}, volume~10, page~S1. BioMed Central,
  2009.

\bibitem{spanier2017radlex}
Assaf~B Spanier, D~Cohen, and Leo Joskowicz.
\newblock A new method for the automatic retrieval of medical cases based on
  the radlex ontology.
\newblock {\em International journal of computer assisted radiology and
  surgery}, 12(3):471--484, 2017.

\end{thebibliography}

\listoffigures

\listoftables

\end{document}